\documentclass[useAMS,usenatbib]{mn2e}
\usepackage{txfonts}
\usepackage{graphicx}
\usepackage{float}
\usepackage{anysize}
\usepackage{times}
\usepackage{epstopdf}
\usepackage{color}

\def\S{\mathcal{S}}

\def\doi{http://doi.org}

\title{Matter density perturbation and power spectrum in running vacuum model}

\author[Chao-Qiang Geng and Chung-Chi Lee]{Chao-Qiang Geng$^{1,2,3}$\thanks{E-mail:
geng@phys.nthu.edu.tw},
Chung-Chi Lee$^{2}$\thanks{E-mail:chungchi@mx.nthu.edu.tw}\\
$^{1}$Chongqing University of Posts \& Telecommunications, Chongqing, 400065, 
China\\
$^{2}$ National Center for Theoretical Sciences, Hsinchu,
Taiwan 300\\
$^{3}$Department of Physics, National Tsing Hua University,
Hsinchu, Taiwan 300}

\begin{document}

\date{}


\maketitle

\label{firstpage}

\begin{abstract}
We investigate the matter density perturbation $\delta_m$ and   power spectrum $P(k)$ in the running vacuum model (RVM)
with the cosmological constant being a function of the Hubble parameter, given by $\Lambda = \Lambda_0 + 6 \sigma H H_0+ 3\nu H^2$,
in which the linear and quadratic terms of $H$ would originate from the QCD vacuum condensation and  cosmological renormalization group, respectively.
Taking the dark energy perturbation into consideration, we derive the evolution equation for $\delta_m$ and find a specific scale 
$d_{cr}=2 \pi/k_{cr}$, which divides the evolution of the universe into the sub and super-interaction regimes,
corresponding to $k \ll k_{cr}$ and $k \gg k_{cr}$, respectively. For the former, the evolution of $\delta_m$ has the same behavior 
as that in the $\Lambda$CDM model, while for the latter, the growth of $\delta_m$ is frozen (greatly enhanced) when $\nu + \sigma >(<)0$ 
 due to the couplings between radiation, matter and dark energy.
It is clear that the observational data rule out the cases with $\nu<0$ and $\nu + \sigma <0$, while the allowed window for the model parameters 
is extremely  narrow with $\nu, |\sigma| \lesssim \mathcal{O}(10^{-7})$.
\end{abstract}


   \maketitle
%

\section{Introduction} \label{sec:introduction}

It is well-known that the Type-Ia supernova observations (Riess et al. \citep{Riess:1998cb}; Perlmutter et al. \citep{Perlmutter:1998np})
have revealed the late-time accelerating expansion of our universe.
To realize the accelerating universe, 
it is necessary to introduce a negative pressure fluid to the gravitational theory, referred to as
``Dark Energy'' (Copeland et al. \citep{Copeland:2006wr}), while the simplest scenario is to have
the cosmological constant $\Lambda$, such as the $\Lambda$CDM model.
Currently, the  $\Lambda$CDM model perfectly fits the observational data, but leaves several difficulties, such as the ``fine-tuning" (Weinberg \citep{Weinberg:1988cp}; Weinberg \citep{WBook}) and ``coincidence'' (Ostriker and Steinhardt \citep{Ostriker:1995rn}; Arkani-Hamed et al. \citep{ArkaniHamed:2000tc}) problems.

In this work, we are interested in
 the running vacuum model (RVM), which has been used to solve the ``coincidence'' problem (Ozer and Taha~\citep{Ozer:1985ws}; Carvalho et al. \citep{Carvalho:1991ut}; Lima and Maia \citep{Lima:1994gi}; Lima and Trodden \citep{Lima:1995ea}; Overduin and Cooperstock \citep{Overduin:1998zv};  Dymnikova and Khlopov \citep{Dymnikova:2001ga}; Carneiro and Lima \citep{Carneiro:2004iz};  Bauer \citep{Bauer:2005rpa}; Shapiro et al. \citep{Shapiro:2004ch}; Alcaniz and Lima\citep{Alcaniz:2005dg}; Barrow and Clifton \citep{Barrow:2006hia};
 Shapiro and Sola \citep{Shapiro:2009dh}; Geng and Lee \citep{Geng:2016epb}; Geng et al. \citep{Geng:2016dqe}).
In this model, the cosmological constant evolves in time and decays to radiation and matter in the evolution of the universe, leading to the same order of magnitude for the energy densities of dark energy and  dark matter.
Its observational applications have  been also extensively explored in the literature (Espana-Bonet et al. \citep{EspanaBonet:2003vk}; Tamayo et al. \citep{Tamayo:2015qla}).
Additionally, it has been shown that the RVM can fit various observational data, indicating that this scenario is good in describing the evolution history of our universe (Sola \citep{Sola:2016vis}; Sola et al. \citep{Sola:2015wwa}; Sola et al. \citep{Sola:2016jky}; Sola et al. \citep{Sola:2016ecz}).
In our study, we will concentrate on
 the specific model with $\Lambda= \sum\limits_{i=0}^{2} \lambda_i H^i$ (Borges and Carneiro \citep{Borges:2005qs}; Borges et al. \citep{Borges:2007bh}; Carneiro et al. \citep{Carneiro:2007bf}; Zimdahl et al. \citep{Zimdahl:2011ae}; Sola \citep{Sola:2013gha}; Sola and Gomez-Valent \citep{Sola:2015rra}), in which the quadratic term, $\lambda_2 H^2$, might come from the quantum effects induced by the cosmological renormalization group (Alcaniz et al. \citep{Alcaniz:2012mh}; Costa et al. \citep{Costa:2012xw}; Sola \citep{Sola:2014tta}; Gomez-Valent et al. \citep{Gomez-Valent:2014rxa}), while the linear term, $\lambda_1 H$, would originate from the theory with the QCD vacuum condensation associated with the chiral phase transition (Schutzhold \citep{Schutzhold:2002pr}; Banerjee et al.\citep{Banerjee:2003fg}; Klinkhamer and Volovik \citep{Klinkhamer:2009nn}; 
 Ohta \citep{Ohta:2010in}; Cai et al. \citep{Cai:2010uf}).

When it comes to the decaying dark energy model, it is reasonable to consider not only the background evolution equations but also the density perturbation of dark energy.
We follow the same method in the references (Fabris et al. \citep{Fabris:2006gt}; Borges et al. \citep{Borges:2008ii}) to rewrite  dark energy as a function of a Lorentz scalar $\nabla_{\mu} U^{\mu}$, where $U^{\mu}=dx^{\mu}/\sqrt{-ds^2}$ is the four-velocity.
Based on such an expression, we examine the matter density perturbation $\delta_m$ and  power spectrum $P(k)$ in the linear perturbation theory of gravity.
Note that in the literature (Fabris et al. \citep{Fabris:2006gt}), the matter density perturbation evolves from $z=1100$ (the recombination era) to $z=0$ (the present), where the initial conditions are taken from the $\Lambda$CDM limit with the BBKS transfer function.
However, the density perturbation of the RVM may influence the evolution of the matter density perturbation in the high redshift regime.
We take the scale invariance initial conditions at the very early time of the universe, in which  all the perturbation modes are at the super-horizon scale with the same behavior  as that in the $\Lambda$CDM model.
Then, we analyze the properties in the 
sub and super-interaction scales with the allowed ranges for the model parameters  discussed.

This paper is organized as follows:
We briefly introduce the running vacuum model in Sec.~\ref{sec:model}. We derive the linear perturbation equations 
with the synchronous gauge and the evolution property of the matter density perturbation in Sec.~\ref{sec:density-pert}.
In Sec.~\ref{sec:mpk}, we show the evolutions of $\delta_m$ and  $P(k)$.
Our conclusions are presented in Sec.~\ref{sec:conclusion}.

\section{Running cosmological constant model}
\label{sec:model}

We start from the Einstein equation with $\kappa^2 = 8 \pi G =1$,
\begin{eqnarray}
\label{eq:fieldeq}
R_{\mu \nu} - \frac{g_{\mu \nu}}{2}R + \Lambda g_{\mu \nu} = T_{\mu \nu}^M \,,
\end{eqnarray}
where $R=g^{\mu \nu} R_{\mu \nu}$ is the Ricci scalar, $\Lambda$ is the time-dependent cosmological constant, and $T_{\mu \nu}^M$ is the energy-momentum tensor of matter and radiation.
In the Friedmann-Lema\"itre-Robertson-Walker (FLRW) metric,
\begin{eqnarray}
\label{eq:pert_metric0}
ds^2 =  -dt^2 + a^2(t) \left[ \delta_{ij} dx^i dx^j \right] \,,
\end{eqnarray}
we obtain,
\begin{eqnarray}
\label{eq:Friedmann-1}
&& H^2= \frac{1}{3} \left( \rho_{M} + \rho_{\Lambda} \right) \,, \\
\label{eq:Friedmann-2}
&& \dot{H} = - \frac{1}{2} \left( \rho_{M}+  P_{M} + \rho_{\Lambda} + P_{\Lambda} \right) \,,
\end{eqnarray}
where $H=\dot{a}/a$ presents the Hubble parameter, $\rho_{M}$ ($P_{M}$) corresponds to the total energy density (pressure) of matter and radiation, and $\rho_{\Lambda}$ ($P_{\Lambda}$) is the energy density (pressure) of the cosmological constant, derived from Eq.~(\ref{eq:fieldeq}),
\begin{eqnarray}
\label{eq:rhode}
\rho_{\Lambda} = -P_{\Lambda} = \Lambda(H)\,.
\end{eqnarray}
In the running cosmological constant model, $ \Lambda(H)$ is taking  to be a function of the Hubble parameter $H$ 
 (Basilakos et al. \citep{Basilakos:2009wi}; Gomez-Valent and Sola \citep{Gomez-Valent:2014fda}; Gomez-Valent et al. \citep{Gomez-Valent:2015pia}), given by
\begin{eqnarray}
\label{eq:rnlam}
\Lambda(H) =  \Lambda_0 + 6 \sigma H_0 (H-H_0) + 3 \nu (H^2-H_0^2) \,,
\end{eqnarray}
where $\nu$, $\sigma$ and $\Lambda_0$ are free parameters, while $H_0$ is the Hubble parameter at the present.
We note that the linear and quadratic teams in Eq.~(\ref{eq:rnlam}) could originate from 
two possible physical sources of
the QCD vacuum condensation associated with the chiral phase transition (Schutzhold \citep{Schutzhold:2002pr}; Borges and Carneiro \citep{Borges:2005qs})
and the quantum effect induced by the cosmological renormalization group running of the vacuum energy in curved space-time (Sola \citep{Sola:2013gha}), respectively.
Substituting Eq.~(\ref{eq:rnlam}) into the conservation equation $\nabla^{\mu} (T^M_{\mu \nu}+T^\Lambda_{\mu \nu}) = 0$, 
we obtain
\begin{eqnarray}
\label{eq:contl}
&&\dot{\rho}_{\Lambda} + 3 H (1+w_{\Lambda}) \rho_{\Lambda} = \dot{\rho}_{\Lambda} \,, \\
\label{eq:contm}
&& \dot{\rho}_m + 3 H \rho_m = -R_m \dot{\rho_{\Lambda}} \,, \\
\label{eq:contr}
&& \dot{\rho}_r + 4 H \rho_r = -R_r \dot{\rho_{\Lambda}} \,,
\end{eqnarray}
where $R_{r(m)}$ represents the interaction rate between radiation (matter) and dark energy with
\begin{eqnarray}
\label{eq:Qmr}
R_{r(m)}=  \frac{\rho_{r(m)}+P_{r(m)}}{\rho_M+P_M} \,,
\end{eqnarray}
respectively, where $\rho_M = \sum\limits_{\ell=r,m} \rho_\ell$ and $P_M = \sum\limits_{\ell=r,m} P_\ell$.
Note that the signs for the model parameters $\nu$ and $\sigma$ 
will be carefully examined.

In order to investigate the dynamics of  dark energy, the cosmological constant in Eq.~(\ref{eq:rnlam}) should be represented as a function of a Lorentz scalar.
By taking the FLRW metric, the covariant derivative of the four-velocity $U^{\mu} \equiv dx^{\mu} / \sqrt{-ds^2} $ is given by
\begin{eqnarray}
\label{eq:hubbleU}
\nabla_{\mu} U^{\mu} = 3H\,.
\end{eqnarray}
As a result,
$\Lambda(H)$ can be rewritten as
\begin{eqnarray}
\label{eq:rnlam2}
\Lambda =  \Lambda_0 -3(2 \sigma +\nu)H_0^2 + 2 \sigma \nabla_{\mu} U^{\mu} + \frac{\nu}{3} (\nabla_{\mu} U^{\mu})^2 \,.
\end{eqnarray}
Even though the expression for the Hubble parameter is not unique, the relation in Eq.~(\ref{eq:hubbleU}) is the simplest 
way to rewrite the Hubble parameter to be a Lorentz scalar (Fabris et al. \citep{Fabris:2006gt}; Borges et al. \citep{Borges:2008ii}; Velasquez-Toribio \citep{VelasquezToribio:2009qp}).

\section{Linear Perturbation Theory}
\label{sec:density-pert}

Since the model with the strong couplings between radiation/matter and $\Lambda$, corresponding to $\nu, \sigma \sim \mathcal{O}(1)$ is unable to fit  the current astrophysical and cosmological observations (Gomez-Valent and Sola \citep{Gomez-Valent:2014fda}; Gomez-Valent et al. \citep{Gomez-Valent:2015pia}), we only focus on the small ones with $\nu, \sigma \ll 1$.
Following the standard procedure (Ma and Bertschinger \citep{Ma:1995ey}), we calculate the evolutions of linear perturbation equations under the synchronous gauge with the perturbed dark energy density.
The metric perturbation is given by
\begin{eqnarray}
\label{eq:pert_metric}
ds^2 = -dt^2 + a^2(t) \left[  (\delta_{ij} + h_{ij}) dx^i dx^j \right] \,,
\end{eqnarray}
with
\begin{eqnarray}
\label{eq:pert_h}
h_{ij} = \int d^3 k e^{i \vec{k} \cdot \vec{x}} \left[ \hat{k}_i \hat{k}_j h(\vec{k},\tau) + 6 \left( \hat{k}_i \hat{k}_j -\frac{1}{3} \delta_{ij} \right) \eta(\vec{k},\tau) \right] \,,
\end{eqnarray}
where $i,j=1,2,3$, $h$ and $\eta$ are two scalars in the synchronous gauge, and $\hat{k} = \vec{k}/ k $ is the k-space unit vector.
 From the relation
\begin{eqnarray}
\label{eq:pertv}
\nabla_{\mu} U^{\mu}=3H + \left( \theta +\frac{\dot{h}}{2} \right)\,,
\end{eqnarray}
the density perturbation of dark energy is given by
\begin{eqnarray}
\label{eq:drholam}
\delta \rho_{\Lambda} = 2 (\sigma H_0 + \nu H) \left( \theta +\frac{\dot{h}}{2} \right) \,,
\end{eqnarray}
where $\theta \equiv \nabla_i \delta U^{i}$ is the momentum perturbation.
From the conservation equation $\nabla^{\mu} (T^m_{\mu \nu} + T^r_{\mu \nu} + T^\Lambda_{\mu \nu}) = 0$ with $\delta
 T^0_0 = \sum\limits_{\ell=r,m} \delta \rho_\ell$, $\delta
 T^0_i = - \delta T^i_0 = \sum\limits_{\ell=r,m} (\rho_\ell + P_\ell) v^i_a$ and $\delta
 T^i_j = \sum\limits_{\ell=r,m} \delta P_\ell \delta^i_j$, the evolutions of 
 the density perturbation $\delta_\ell \equiv \delta\rho_\ell/\rho_\ell$ and  momentum perturbation $\theta_\ell \equiv \partial_i v^i_\ell = \theta \rho_\ell/ \rho_M$ 
 ($\ell=r$ or $m$) can be derived. 
 Explicitly, one gets 
\begin{eqnarray}
\label{eq:drho}
&&\dot{\delta}_{\ell} = - \left(1+w_{\ell} \right)\left( \theta_{\ell} + \frac{\dot{h}}{2} \right) \nonumber \\
&& \qquad + R_{\ell} \left[ \frac{\dot{\rho}_{\Lambda}}{\rho_{\ell}} \delta_{\ell} \ - \frac{\left( \rho_\Lambda \dot{\delta}_{\Lambda} + \dot{\rho}_{\Lambda} \delta_{\Lambda} \right)}{\rho_{\ell}} \right] \,, \\
 \label{eq:dtheta}
&&\dot{\theta}_{\ell} = -H\left( 1-3w_{\ell} \right) \theta_{\ell} + \frac{w_{\ell}}{1+w_{\ell}} \frac{k^2}{a^2} \delta_{\ell} \nonumber \\
&& \qquad + R_{\ell} \left(  \frac{\dot{\rho}_{\Lambda}}{\rho_{\ell}} \theta_{\ell} - \frac{k^2}{a^2} \frac{\rho_\Lambda \delta_{\Lambda}}{\rho_{\ell}} \right) \,,
\end{eqnarray}
where $w_\ell \equiv P_\ell / \rho_\ell = \delta P_\ell / \delta \rho_\ell$ and $\dot{w}_\ell = 0$ are used in Eqs.~(\ref{eq:drho}) and (\ref{eq:dtheta}).
In addition, the evolution of the synchronous scalar $h(a,k)$ is given by,
\begin{eqnarray}
\label{eq:hdot}
\ddot{h} + 2H \dot{h}= -\sum\limits_{\ell=r,m} \left( 1+3w_\ell \right) \delta \rho_\ell \,
\end{eqnarray}
from the field equation in Eq.~(\ref{eq:fieldeq}).

In the matter dominated era, as the radiation density is sub-dominated to the universe, $i.e.$ $\rho_r \ll \rho_m$, 
from Eqs.~(\ref{eq:drholam})-(\ref{eq:hdot}), we find that
\begin{eqnarray}
\label{eq:drho2}
\delta_m^{\prime} \equiv \frac{d \delta_m}{d N}  \simeq  - \left[ 1-\frac{4}{9} \frac{k^2}{a^2 H^2} \left( \nu + \sigma \frac{H_0}{H} \right) \right] \left( \frac{\theta_m}{H} + \frac{h^{\prime}}{2} \right) \nonumber \\
- \left(2\nu- \sigma\frac{H_0}{H} \right) \delta_m \,,
\end{eqnarray}
\begin{eqnarray}
\label{eq:dtheta2}
\theta_m^{\prime} \equiv \frac{d \theta_m}{d N} \simeq  -\theta_m - \frac{2}{3}\frac{k^2}{a^2 H} \left( \nu + \sigma \frac{H_0}{H} \right) \left( \frac{\theta_m}{H} + \frac{h^{\prime}}{2} \right) \,,
\end{eqnarray}
where $h^\prime \equiv d h/d N$ and $N=\ln a$ with the higher order terms of $\nu$ and $\sigma$  neglected.
Combining Eqs.~(\ref{eq:hdot})-(\ref{eq:dtheta2}), we obtain the second order derivative equation of the matter density perturbation
to be
\begin{eqnarray}
\label{eq:ddrho}
\delta_m^{\prime \prime} + \left[ \frac{1}{2} + \frac{2}{3}\frac{ \tilde{k}^2}{a^2 H^2}  \right] \delta_m^{\prime} - \left[ \frac{3}{2} -  \frac{2\nu \tilde{k}^2}{a^2 H^2}  \right] \delta_m \simeq 0\,,
\end{eqnarray}
where
\begin{eqnarray}
\tilde{k}^2= \left( \nu + \sigma \frac{H_0}{H} \right) k^2 \,.
\end{eqnarray}

\begin{figure}
\centering
\includegraphics[width=1.0 \linewidth]{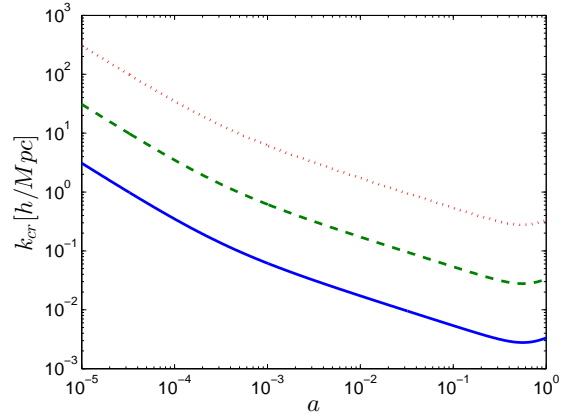}
\caption{The critical scale factor $a_{cr}$ as a function of the wavenumber $k$ with $\sigma=0$ and $\nu=10^{-2}$ (solid line), $10^{-4}$ (dashed line) and $10^{-6}$ (dotted line), where the boundary conditions of $\Omega_m=0.26$ and $\Omega_r=8.4\times 10^{-5}$ are used.}
\label{fg:1}
\end{figure}
From Eq.~(\ref{eq:ddrho}), we see that the behaviors of the matter density perturbation
 at the sub and super-interaction scales, corresponding to
 $\vert \tilde{k}^2 \vert /a^2 \ll H^2$ and $\vert \tilde{k}^2 \vert /a^2 \gg H^2$, respectively,
 are quite different.
At the sub-interaction scale, Eq.~(\ref{eq:ddrho}) reduces to the $\Lambda$CDM case,
\begin{eqnarray}
\label{eq:ddrho-lcdm}
\delta_m^{\prime \prime} + \frac{1}{2} \delta_m^{\prime} - \frac{3}{2} \delta_m = 0\,,
\end{eqnarray}
and the growth of $\delta_m$ increases during the expansion of the universe, $\delta_m \propto a$.
On the other hand,
when the perturbation mode $\delta(k,a)$ enters the super-interaction regime, Eq.~(\ref{eq:ddrho}) becomes
\begin{eqnarray}
\label{eq:ddrho-rn}
\delta_m^{\prime \prime} + \frac{2}{3}\frac{\tilde{k}^2}{a^2 H^2}  \delta_m^{\prime} + \frac{2\nu \tilde{k}^2}{a^2 H^2} \delta_m = 0\,,
\end{eqnarray}
which implies that $\delta_m$ is suppressed (enhanced) by the dark energy perturbation 
with $\left( \nu + \sigma H_0/H \right) > 0$ ($<0$).
As a result, we can define the super-interaction divide $k_{cr} = 2 \pi/ d_{cr}$ at $\tilde{k}^2_{cr} = a^2 H^2$,
given by
\begin{eqnarray}
\label{eq:kcr}
k_{cr}^2 = \frac{a^2 H^2}{\vert \nu + \sigma  H_0/H \vert}\,.
\end{eqnarray}

In Fig.~\ref{fg:1}, we plot the super-interaction divide $k_{cr}$ as a function of the scale factor $a$ with $\sigma=0$, $\Omega_m=0.26$ and $\Omega_r=8.4 \times 10^{-5}$. We see that
$k_{cr}$ decreases in time and reaches the minimum at $a_M = \left[ \Omega_m/2(1-\Omega_m) \right]^{1/3} \simeq 0.56$.
The super-interaction effect appears in a very small region (large $k$) in the very early time of the universe.
In addition, the larger $\nu$ is, the wider influences of  dark energy  to the matter density perturbation will be.
Consequently, the size of the super-interaction regime enlarges in the cosmological evolution and finally reaches the scale  $k_{cf}^M \simeq 2.8 \times 10^{-3}$, $2.8 \times 10^{-2}$ and $2.8 \times 10^{-1} [h/Mpc]$ at $a=a_M$ for $\nu=10^{-2}$, $10^{-4}$ and $10^{-6}$, respectively.

\section{Evolutions of  Matter Density Perturbation and  Power Spectrum}
\label{sec:mpk}

The Hubble radius $d_H \equiv H^{-1}$ increases during the evolution of our universe.
More and more density perturbation modes of $\delta_k$ with the wavenumber $k$ enter the horizon, 
in which $k^2 = a^2 H^2$ with $a=a_k$.
Explicitly, using $\Omega_m=0.26$, $\Omega_r=8.4 \times 10^{-5}$ and $H_0=70 km/s \cdot Mpc$, 
we find that 
$k=10^{-3}$ and $0.25 [h/Mpc]$ go to the horizon at $a_k \simeq 0.03$ and $1.24 \times 10^{-5}$, respectively.
Following the similar procedure in the literature (Borges et al. \citep{Borges:2007bh}), we can recover $P(k)$ in the $\Lambda$CDM model with the BBKS transfer function within $10\%$ accuracy by taking the initial density perturbation to be the scale invariance when all modes are located at the super-horizon scale, i.e., $\delta_m = 3 \delta_r /4 \propto k^{n_s/2}$ at $a \ll a_k$, where $n_s$ is the spectral index, given from the inflationary epoch. From the above discussion, it is reasonable to choose $\ln a=-18$ to be the initial time in our work.
We note that from Eq.~(\ref{eq:kcr}), the super-interaction effect appears at the scale deep inside the horizon $d_{cr} = \sqrt{\vert \nu + \sigma  H_0/H \vert} \cdot d_H$, indicating that the $\Lambda$CDM initial condition is applicable to the RVM.
In this section, we analyze the evolution of $\delta_m$ from Eq.~(\ref{eq:ddrho-rn}) for two cases:  (i) $\sigma=0$ and (ii) $\sigma \neq 0$, and numerically solve $\delta_m(a)$ and $P(k)$.

\vspace{0.5em}%
\noindent {\em (i) $\sigma = 0$}:

\begin{figure}
\centering
\includegraphics[width=1.0 \linewidth]{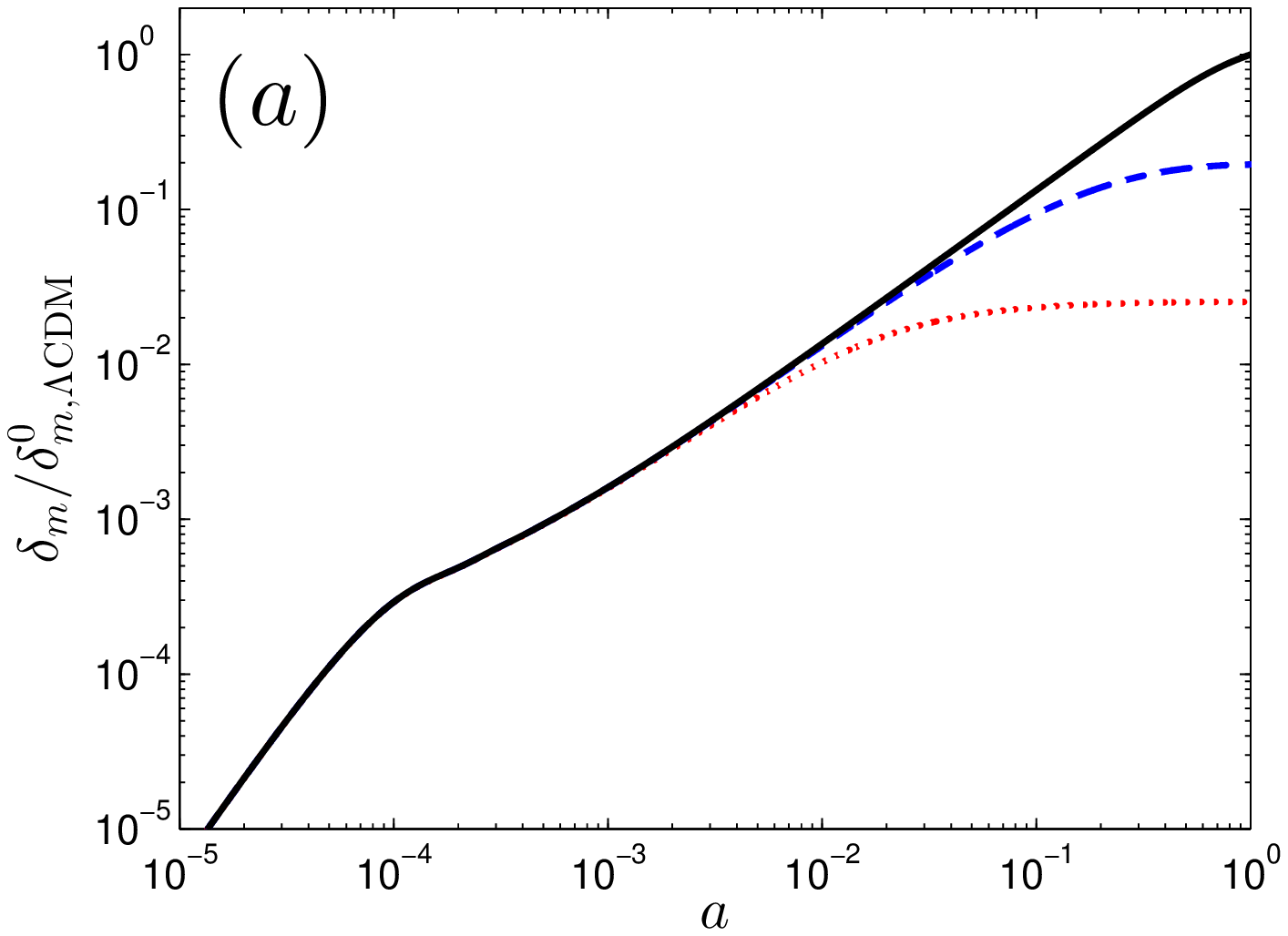}
\includegraphics[width=1.0 \linewidth]{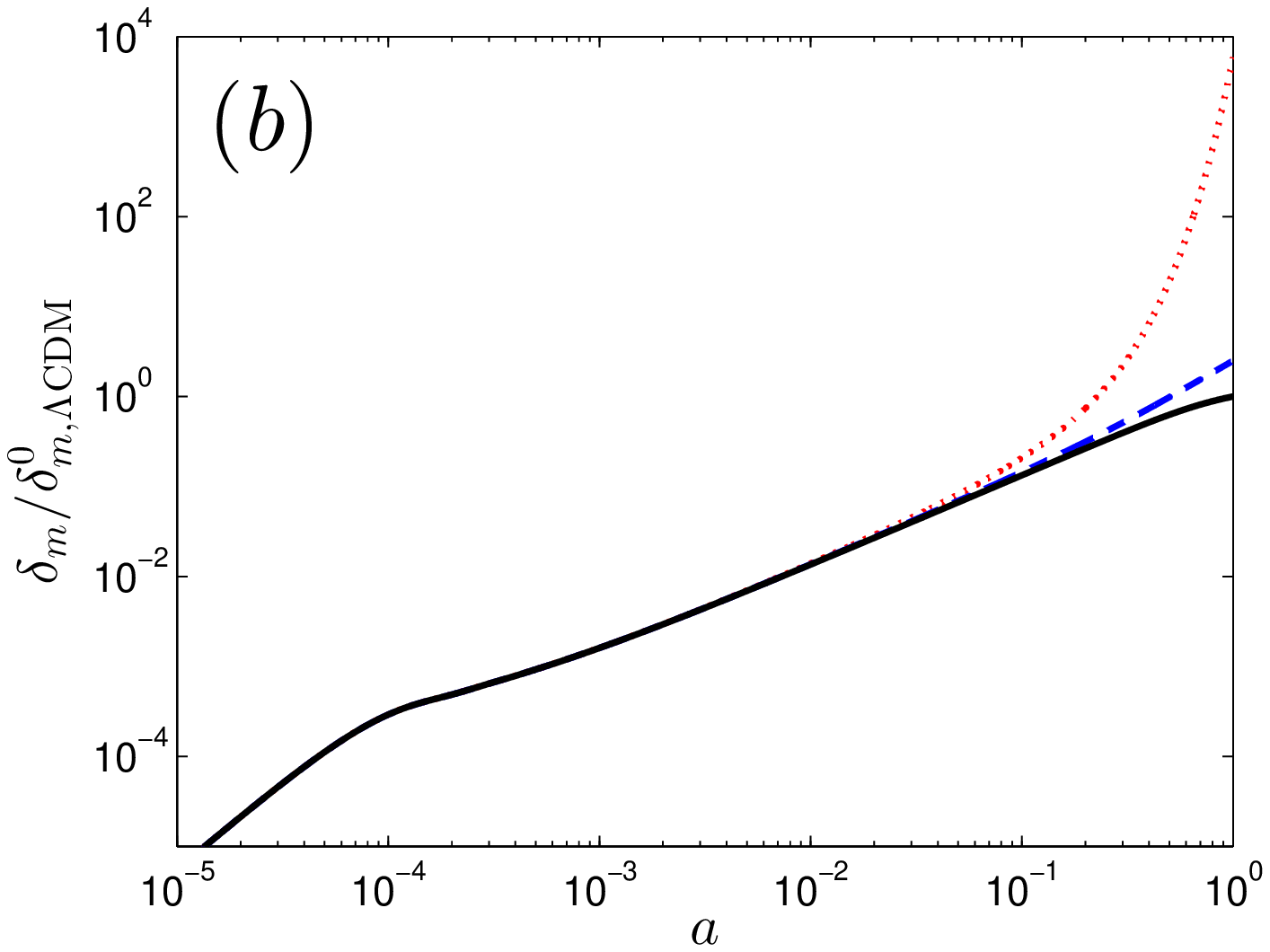}
\caption{Evolution of the matter density perturbation $\delta_m$ as a function of the scale factor $a$ with $\sigma=0$ and (a) $\nu=0$ (solid line), $10^{-5}$ (dashed line) and $10^{-4}$ (dotted line), and (b) $\nu=0$ (solid line), $-2 \times 10^{-6}$ (dashed line) and $-10^{-5}$ (dotted line), respectively,
where $\delta^0_{m, \Lambda \mathrm{CDM}}$ is the matter density perturbation with $\nu=\sigma=0$, i.e. the $\Lambda$CDM limit, at $z=0$
and the boundary conditions are taken to be the same as Fig.~\ref{fg:1} with $k=0.25~[h/Mpc]$.}
\label{fg:2}
\end{figure}

At the super-interaction regime ($k \gg k_{cr}$), the friction in Eq.~(\ref{eq:ddrho-rn}) plays 
the most important role in the evolution history of $\delta_m$.
Thus, with $0 < \nu \ll 1$, the growth of $\delta_m$ is frozen with $\delta_m^{\prime \prime} \approx \delta_m^{\prime} \approx 0$, and the matter power spectrum is suppressed by the dark energy perturbation.
From Eqs.~(\ref{eq:drho}) and (\ref{eq:dtheta}), one can derive that $\dot{h} \simeq -2\dot{\delta}_m  <0$ and $\theta_m \simeq 0$ at $k \ll k_{cr}$, showing that $\delta \rho_{\Lambda} = \rho_{\Lambda} \delta_{\Lambda} < 0 $.
When  entering the super-interaction scale, the second term in the RHS of Eq.~(\ref{eq:dtheta2}) dominates, 
and the momentum perturbation of $\theta_m$ becomes the same order of $ \vert \dot{h}/2 \vert$, implying that the RVM heats up the cold dark matter for a large value of  $k$.
As a result, the particles propagate inside the super-interaction divide, and the growth of $\delta_m$ is frozen.

In Fig.~\ref{fg:2}a, we show the evolution of $\delta_m/\delta^0_{m, \Lambda \mathrm{CDM}}$ as a function of 
the scale factor $a$ with $k=0.25~[h/Mpc]$ and
 $\nu=0$ (solid line), $10^{-5}$ (dashed line) and $10^{-4}$ (dotted line), 
where $\delta^0_{m, \Lambda \mathrm{CDM}}$ is the matter density perturbation with $\nu=\sigma=0$, i.e. the $\Lambda$CDM limit, at $z=0$.
Compared to the $\Lambda$CDM case, the frozen behavior of $\delta_m$ occurs at $a_{cr} \sim 1.5\times 10^{-2}$ ($\nu=10^{-6}$) and $4.8 \times 10^{-3}$ ($\nu=10^{-4}$), respectively.
On the other hand, the friction term in Eq.~(\ref{eq:ddrho-rn}) turns into negative if $\nu<0$, and the growth of the
matter density perturbation sharply  increases.
In Fig.~\ref{fg:2}b, we plot  $\delta_m/\delta^0_{m, \Lambda \mathrm{CDM}}$  with $\nu=0$ (solid line), $2 \times 10^{-6}$ (dashed line) and $10^{-5}$ (dotted line).
From the figure, we see that the matter density perturbation $\delta_m$ deviates from that in $\Lambda$CDM  around $a_{cr} \simeq 3.4 \times 10^{-2}$ ($\nu=2\times 10^{-6}$) and $1.5\times 10^{-2}$ ($\nu=10^{-5}$).
One can observe that the RVM scenario with $\nu<0$ is extremely different from that with $\nu>0$.
Moreover, since the RHS of Eq.~(\ref{eq:dtheta2}) is negative,  the momentum perturbation has no lower bound, causing $\theta_m \rightarrow -\infty$.
Consequently,  the super-interaction scale in both positive and negative $\nu$ cases results in 
the consistent outcomes as those in Fig.~\ref{fg:1}.

\begin{figure}
\centering
\includegraphics[width=1.0 \linewidth]{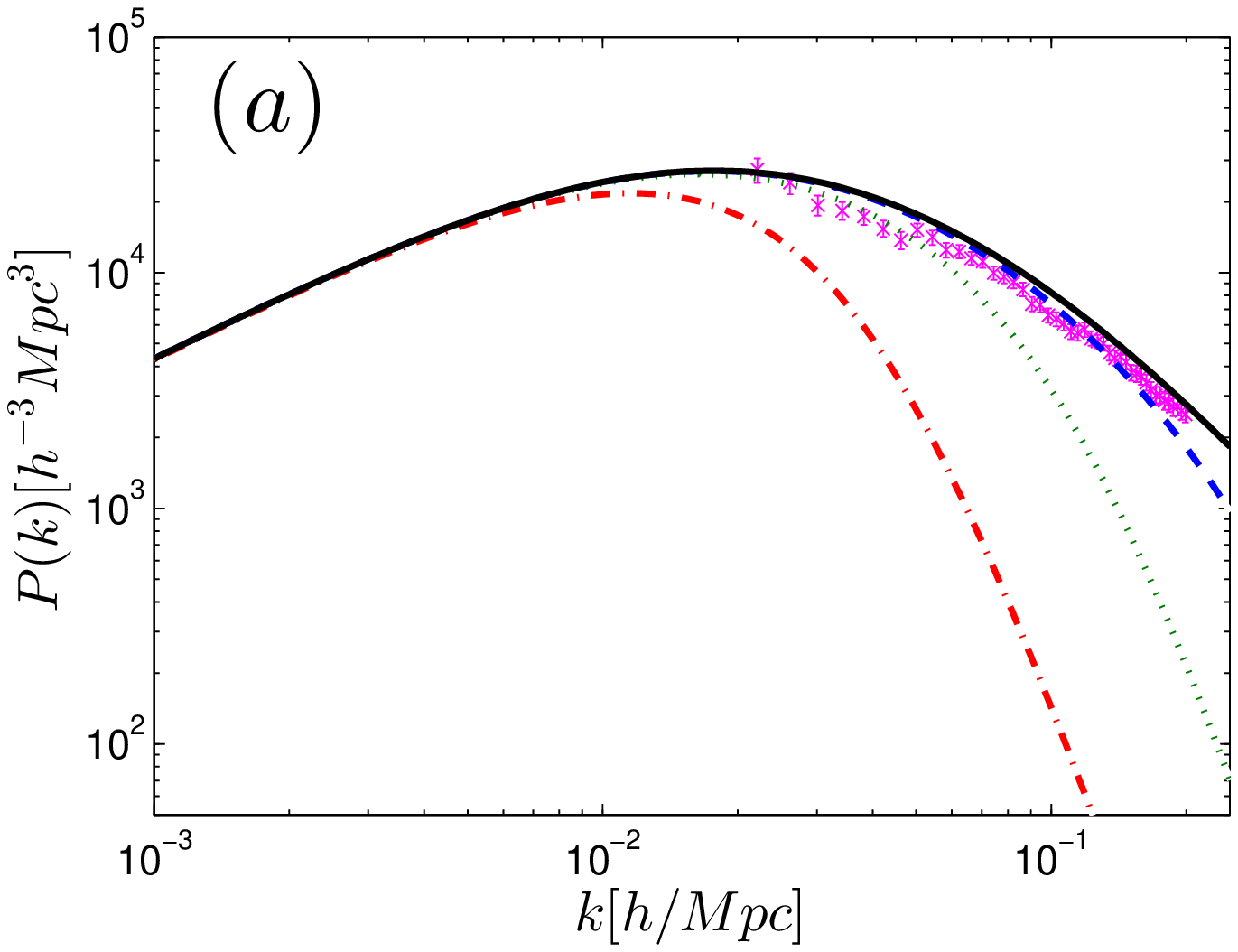}
\includegraphics[width=1.0 \linewidth]{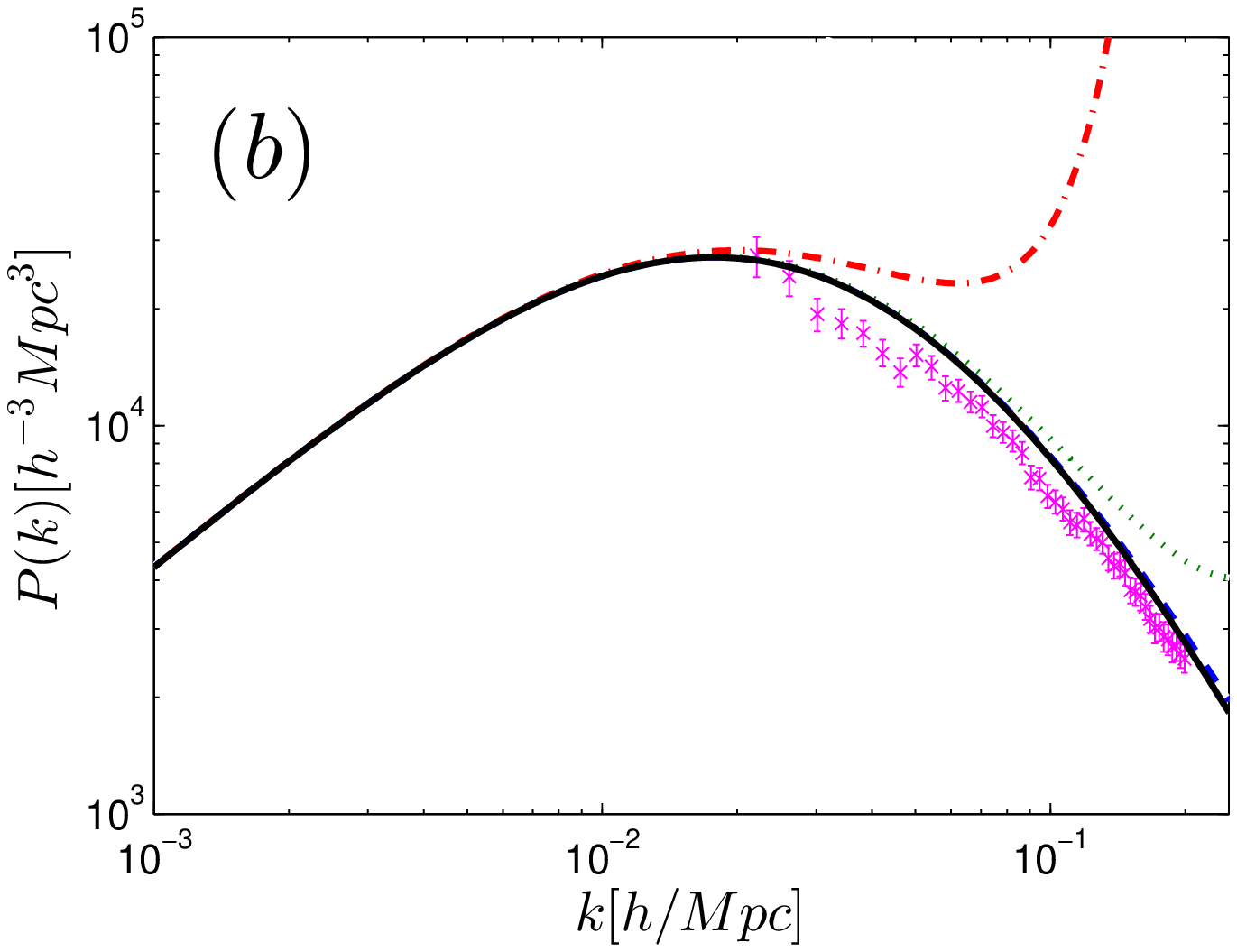}
\caption{ The matter power spectrum $P(k)$ as a function of the wavelength $k$ with $\sigma=0$ and 
(a) $\nu=0$ (solid line), $10^{-6}$ (dashed line), $10^{-5}$ (dotted line) and $10^{-4}$ (dash-dotted line), 
and 
(b) $\nu=0$ (solid line), $-10^{-7}$ (dashed line), $-10^{-6}$ (dotted line) 
and $-10^{-5}$ (dash-dotted line), together with the SDSS LRG DR7 data points, where the boundary conditions are taken to be 
the same as Fig.~\ref{fg:1}.}
\label{fg:3}
\end{figure}
In Fig.~\ref{fg:3}a, we show the matter power spectrum $P(k)$ as a function of the wavenumber $k [h/Mpc]$ with
 $\nu=0$ (solid line), $10^{-6}$ (dashed line), $10^{-5}$ (dotted line) and $10^{-4}$ (dash-dotted line).
Our results in the figure are similar to those in the literature (Fabris et al. \citep{Fabris:2006gt}). The data points in
Fig.~\ref{fg:3}
come from the SDSS LRG DR7  (Reid et al. \cite{Reid:2009xm}).
In this plot, we observe that the suppressed behaviors  for $\nu=10^{-4}$, $10^{-5}$ and $10^{-6}$ become important at the order of $k \gtrsim 0.02$, $0.06$ and $0.2 [h/Mpc]$, respectively, which also support our results in Fig.~\ref{fg:1}.
Clearly, the larger $k$ is, the earlier the $\delta_k$ mode enters the super-interaction regime.

In Fig.~\ref{fg:3}b, we demonstrate the matter power spectrum $P(k)$ as a function of the wavenumber $k$ with 
 $\nu=0$ (solid line), $-10^{-7}$ (dashed line), $-10^{-6}$ (dotted line) and $-10^{-5}$ (dash-dotted line).
As discussed early in this section, when the scale enters the super-interaction regime with $\nu<0$, the evolution of $\delta_m$ acts 
as an increasing mode, leading to the divergence of $P(k)$ at $k \rightarrow \infty$.
This phenomenon clearly illustrates that the negative $\nu$ case fails in describing the evolution of our universe.

\vspace{0.5em}%
\noindent {\em (ii) $\sigma \neq 0$}:

\begin{figure}
\centering
\includegraphics[width=1.0 \linewidth]{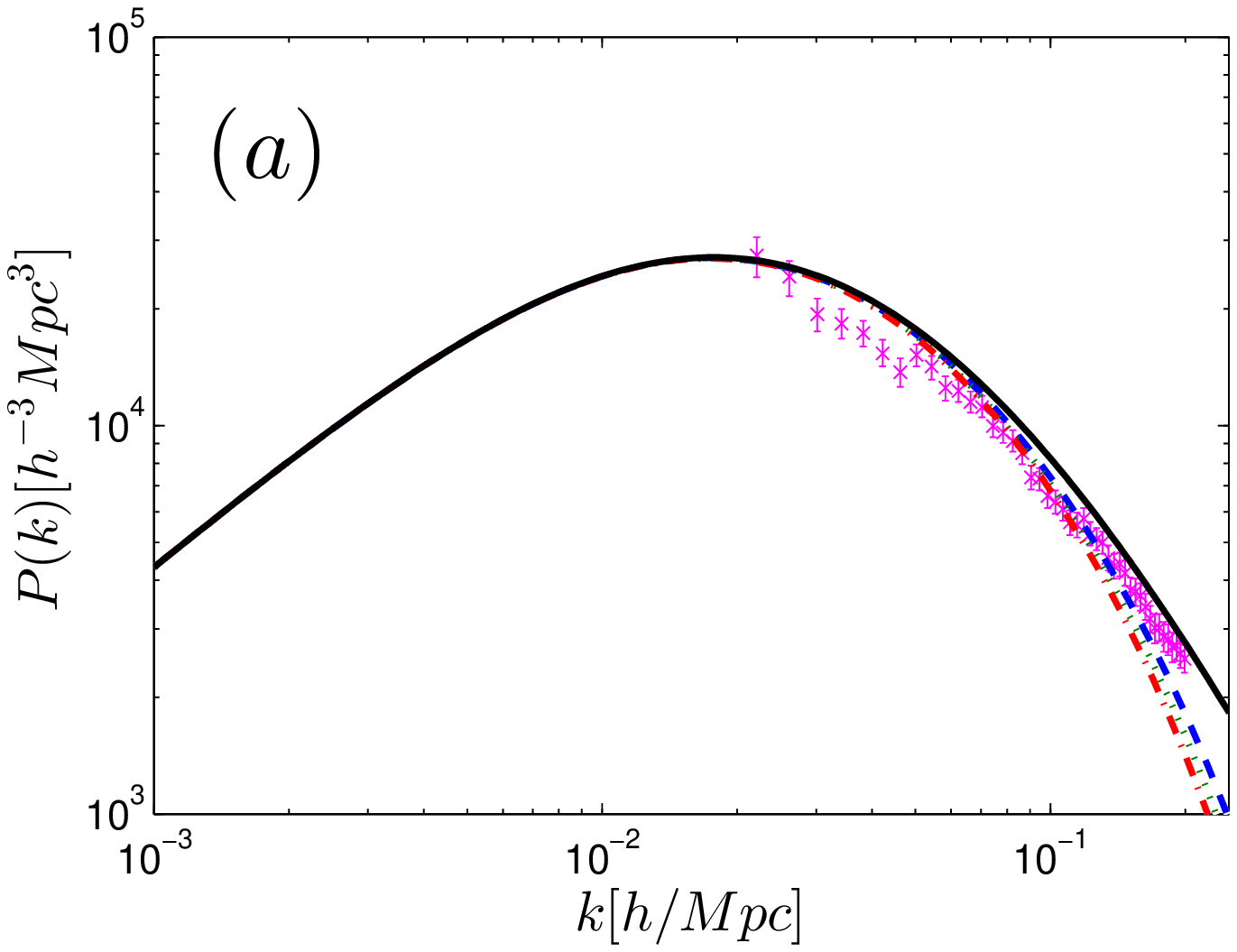}
\includegraphics[width=1.0 \linewidth]{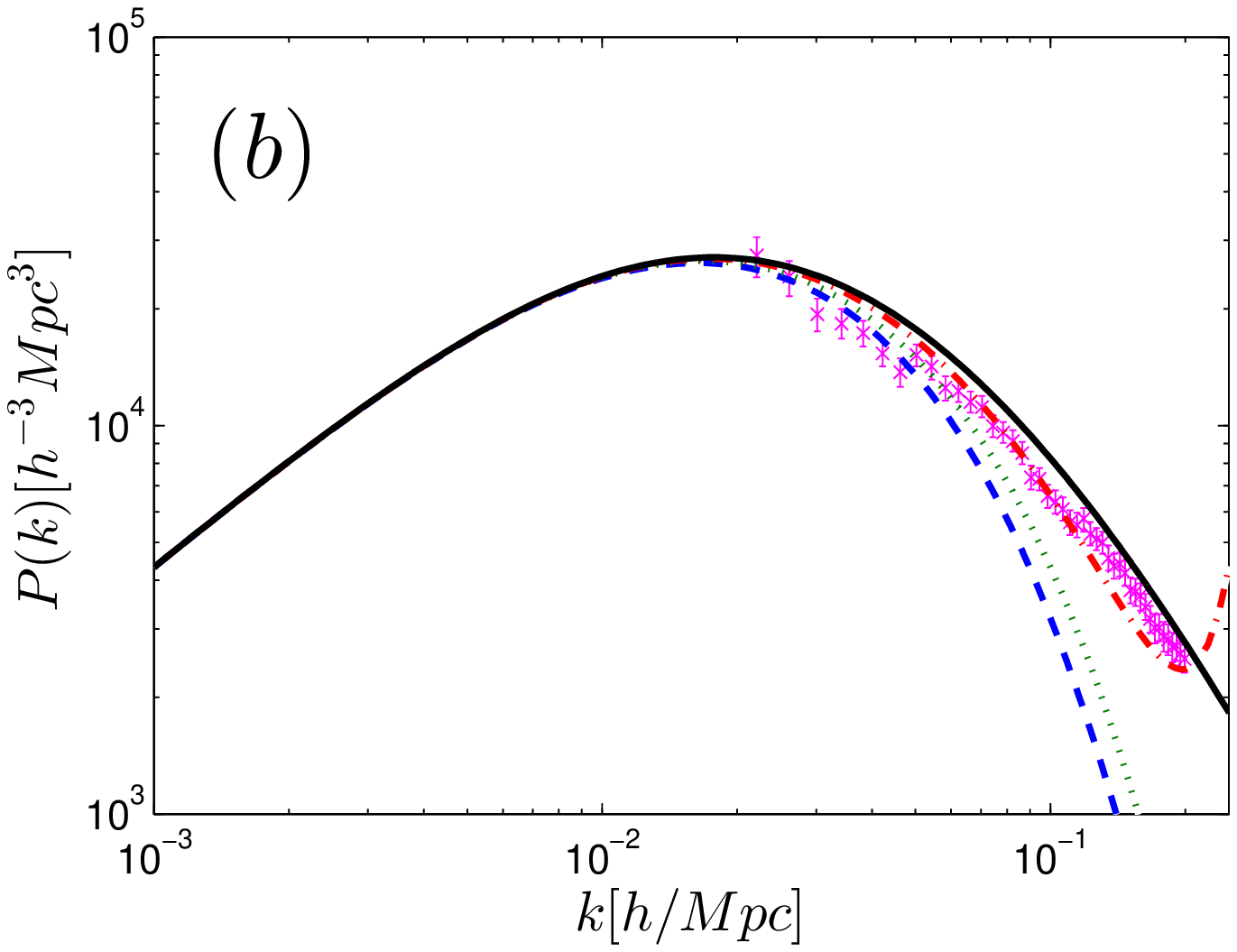}
\caption{The matter power spectrum $P(k)$ as a function of the wavelength $k$ with (a) $(\nu,\sigma)=(0,0)$ (solid line) $(10^{-6},0)$ (dashed line), $(10^{-6},10^{-6})$ (dotted line), $(10^{-6},2 \times 10^{-6})$ (dash-dotted line) and (b) $(\nu,\sigma)=(0,0)$ (solid line) $(10^{-5},0)$ (dashed line), $(10^{-5},-10^{-5})$ (dotted line), $(10^{-5},-2 \times 10^{-5})$ (dash-dotted line).
Legend is the same as Fig.~\ref{fg:3}.}
\label{fg:5}
\end{figure}
In Fig.~\ref{fg:5}a, we display the matter power spectrum $P(k)$ with 
 $(\nu,\sigma)$ = $(0,0)$ (solid line), $(10^{-6},0)$ (dashed line), $(10^{-6},10^{-6})$ (dotted line) and $(10^{-6},2 \times 10^{-6}$ (dash-dotted line).
In Eq.~(\ref{eq:ddrho-rn}), the suppression or enhancement of the matter density perturbation inside the super-interaction regime is controlled by the factor $\nu+\sigma H_0/H$. 
Accordingly, it is  expected that the suppression of $\delta_m$ is strengthened when $\sigma>0$, while $P(k)$ is further suppressed.

In Fig.~\ref{fg:5}b, we exhibit $P(k)$ with  $(\nu,\sigma)$ = $(0,0)$ (solid line), $(10^{-5},0)$ (dashed line), $(10^{-5},-10^{-5})$ (dotted line) and $(10^{-5},-2 \times 10^{-5})$ (dash-dotted line).
This figure shows that the suppression of $P(k)$ is alleviated with $\sigma<0$ at the late-time of the universe.
If $\nu +\sigma H_0/H \rightarrow 0$, the frozen mode of the matter density perturbation melts, and $\delta_m$ starts growing.
However, when $\nu+\sigma H_0 /H < 0$,
the growth of $\delta_m$ turns into the extreme enhancement mode, resulting in the divergence 
of $P(k)$ at $k \rightarrow \infty$.
As a result, we conclude that the RVM in Eq.~(\ref{eq:rnlam2}) with $\nu+\sigma<0$ should  be also ruled out by the instability problem.
\begin{figure}
\centering
\includegraphics[width=1.0 \linewidth]{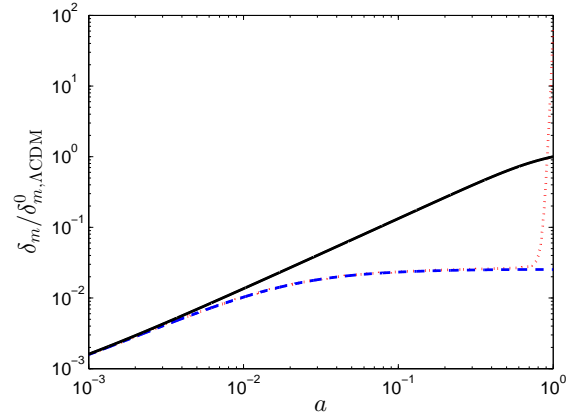}
\caption{Evolution of the matter density perturbation $\delta_m$ as a function of the scale factor $a$ with $(\nu,\sigma)=(0,0)$ (solid line), $(10^{-4},0)$ (dashed line) and $(10^{-4},-1.4 \times 10^{-4})$ (dotted line).
Legend is the same as Fig.~\ref{fg:1}.}
\label{fg:6}
\end{figure}

In Fig.~\ref{fg:6}, we demonstrate the evolution of the density perturbation as a function of the scale factor $a$ with $(\nu,\sigma)=(0,0)$ (solid line), $(10^{-4},0)$ (dashed line) and $(10^{-4},-1.4 \times 10^{-4})$ (dotted line).
With $\Omega_m=0.26$ and $(\nu,\sigma)=(10^{-4},-1.4 \times 10^{-4})$, the turning point of the frozen mode to the increasing 
one can be estimated from the relation,
\begin{eqnarray}
\nu+\frac{\sigma H_0}{H} = 0 \quad \Rightarrow \quad a = \left[ \frac{\Omega_m}{\left( \sigma / \nu \right)^2 + \Omega_m -1} \right] \simeq 0.60 \,,
\end{eqnarray}
which is compatible to the result in Fig.~\ref{fg:6}.

\begin{figure}
\centering
\includegraphics[width=1.0 \linewidth]{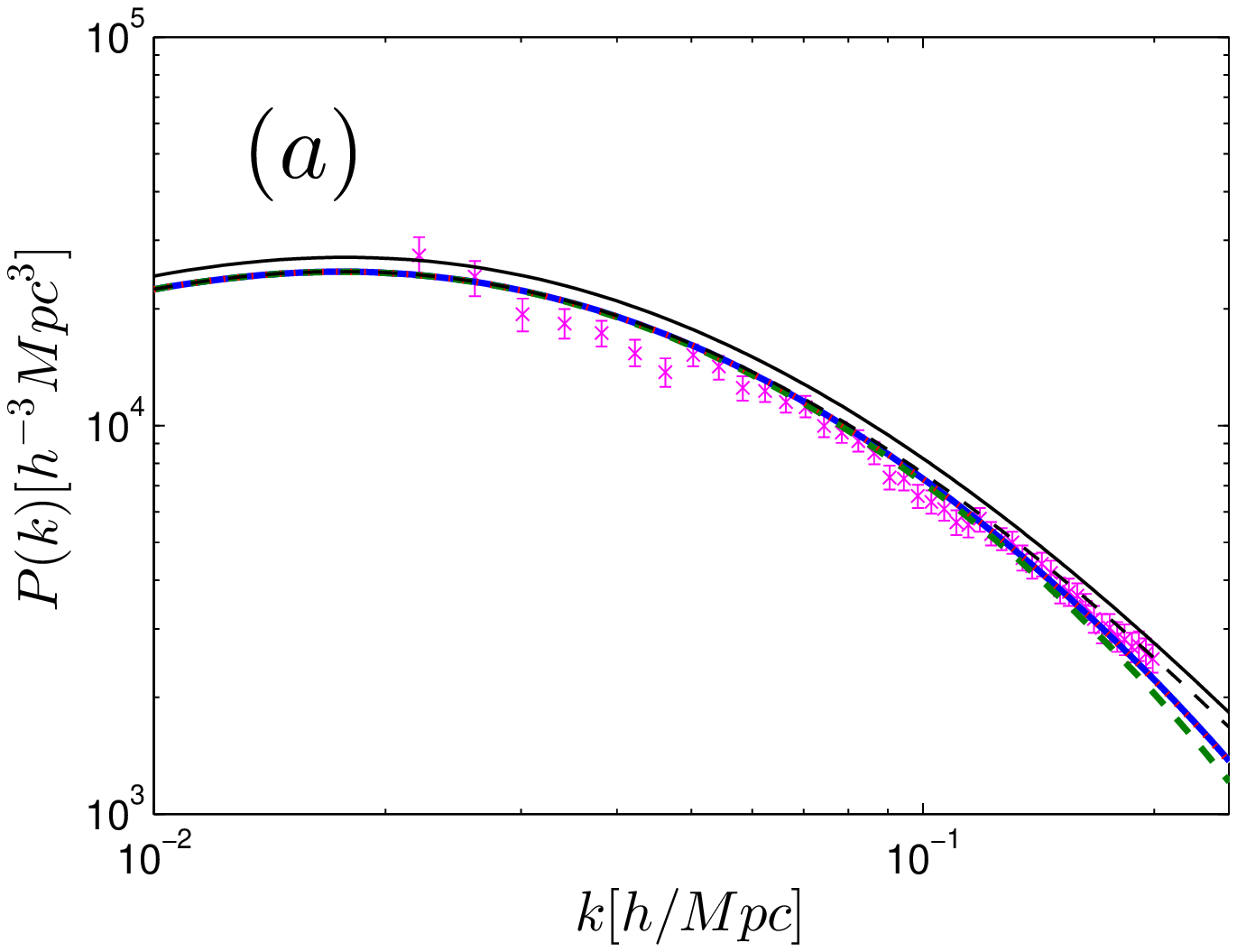}
\includegraphics[width=1.0 \linewidth]{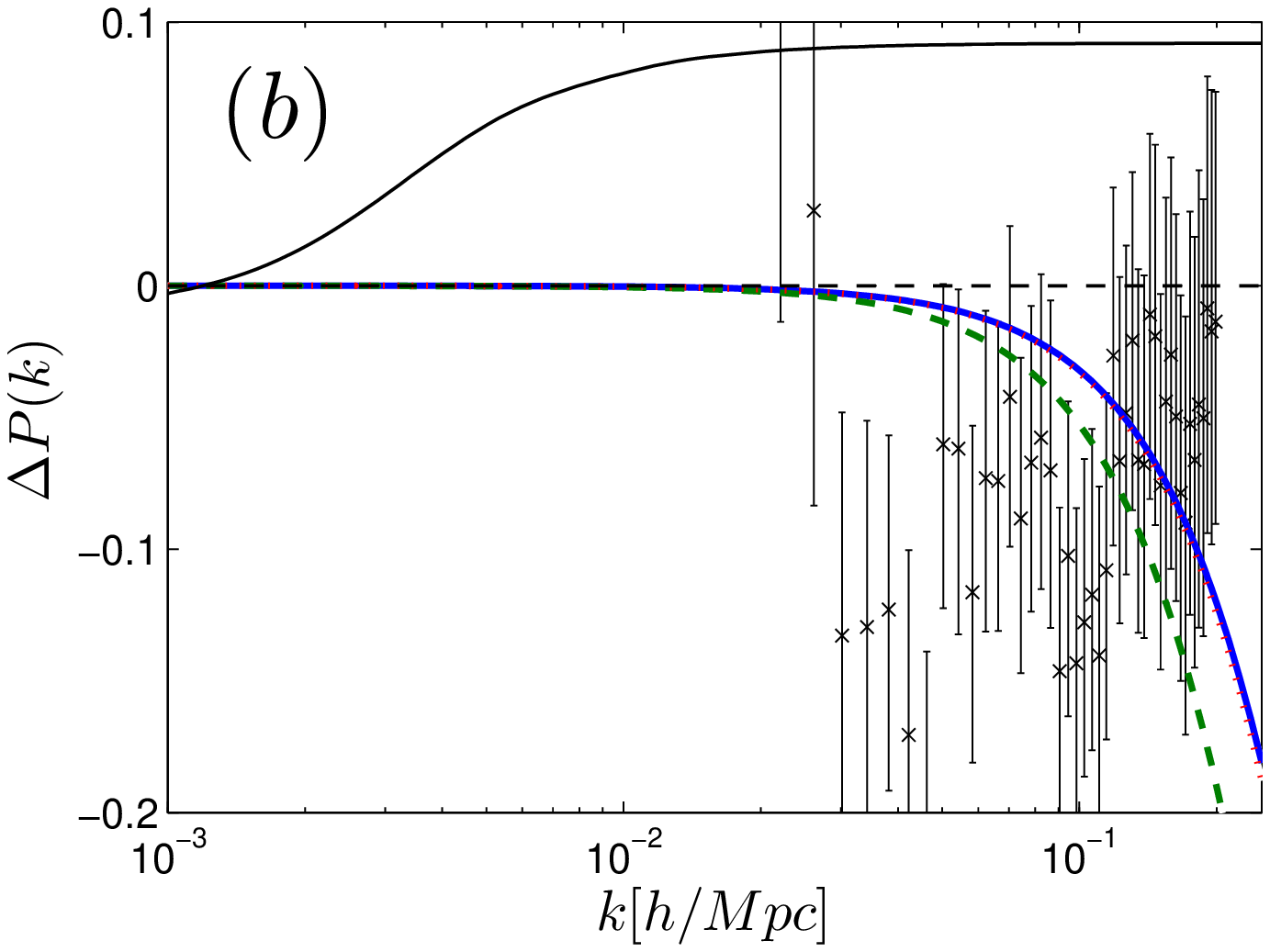}
\caption{(a) The matter power spectrum $P(k)$ and (b) $\Delta P(k) = (P-P_{\Lambda \mathrm{CDM}})/P_{\Lambda \mathrm{CDM}}$ as a function of the wavelength $k$ with selected parameter sets $(\sum m_{\nu}/\mathrm{eV}, \nu, \sigma)= (0, 0, 0)$ (thin solid line), $(0.2, 0, 0)$ (thin dashed line), $(0.2, 3 \times 10^{-7}, 0)$ (thick solid line), $(0.2, 5\times 10^{-7}, 0)$ (thick dashed line) and $(0.2, 5\times 10^{-7}, -5\times 10^{-7})$ (thick dotted line), where $P_{\Lambda \mathrm{CDM}}$ corresponds to the case with $\sum m_{\nu}=0.2$~eV.
Legend is the same as Fig.~\ref{fg:3}.}
\label{fg:7}
\end{figure}

Finally, we take the massive neutrinos into consideration with
\begin{eqnarray}
\Omega_{\nu}h^2= \frac{\sum m_{\nu}}{94~\mathrm{eV}}\,.
\end{eqnarray}
In Fig.~\ref{fg:7}, we plot the matter power spectrum as a function of the
wavenumber $k$ with selected parameter sets $(\sum m_{\nu}/\mathrm{eV}, \nu, \sigma)= (0, 0, 0)$ (thin solid line), $(0.2, 0, 0)$ (thick solid line), $(0.2, 3 \times 10^{-7}, 0)$ (thick dashed line), $(0.2, 5\times 10^{-7}, 0)$ (thick dashed line) and $(0.2, 5\times 10^{-7}, -5\times 10^{-7})$ (thick dotted line).
The matter power spectrum $P(k)$ is suppressed by the free-streaming massive neutrinos in the sub-horizon scale and further frozen due to the decaying dark energy process in the super-interaction regime.
As we can see, the values of $P(k)$ are overlapped for
  $(\nu, \sigma)= (5\times 10^{-7}, -5\times 10^{-7})$ 
  and $(3 \times 10^{-7}, 0)$.
The negative $\sigma$ can alleviate the suppression of $P(k)$ from dark energy, but the effect is limited.
In order to fit the observational data,  we claim that the allowed window for model parameters should be tiny 
with $\nu$, $\vert \sigma \vert \lesssim \mathcal{O}(10^{-7})$.

\section{Conclusions}
\label{sec:conclusion}

We have studied the matter density perturbation $\delta_m$ and  matter power spectrum $P(k)$ in the RVM
with $\Lambda = \Lambda_0 + 6 \sigma H H_0+ 3\nu H^2$.
By rewriting $\Lambda$ to be a function of the covariant derivative of four-velocity as $\Lambda =  \Lambda_0+ 2 \sigma \nabla_{\mu} U^{\mu} + \nu (\nabla_{\mu} U^{\mu})^2/3$, we explicitly derive the linear perturbation equations for  matter and radiation.
The dark energy perturbation $\delta_{\Lambda}$ can be expressed by $\theta$ and $h$, indicating that $\delta_{\Lambda}$ directly couples to $\delta_m$ and $\theta_m$.
We have shown that the growth of $\delta_m$ can be separated into the sub and super-interaction regimes of $\vert \tilde{k}^2 \vert /a^2 \ll H^2$ and $\vert \tilde{k}^2 \vert /a^2 \gg H^2$, respectively.
In the former regime, the interactions between  dark energy and matter are sub-dominated to the evolutions of $\delta_m$ and $\theta_m$, and the growth of $\delta_m$ behaves the same as that in the $\Lambda$CDM model. 
In the later one, the decaying $\Lambda$ drags the evolution of $\delta_m$, and $P(k)$ is suspended (sharply increased) when $\nu+\sigma H_0 / H >(<)\,0$.
The RVM with $\nu<0$ or $\nu+\sigma <0$ is clearly ruled out by the divergences of the physical quantities, $\delta_m$ and $\theta_m$.
We have also found that the model parameters are strongly constrained to be  $\nu>0$ and $\nu+\sigma >0$ with
 $\nu$ and $|\sigma| \lesssim \mathcal{O}(10^{-7})$.

The perturbed RVM modifies not only the growth of $\delta_m$ but the evolution of $\theta_m$.
In the $\Lambda$CDM model, the cold dark matter rests on the comoving frame, i.e., $\theta_m \rightarrow 0$, 
but the behavior of $\theta_m$ in the RVM scenario is totally different.
In the super-interacting regime, $\delta_m$ is frozen, but $\theta_m$ is enhanced to be a non-zero value, 
indicating that the massive cold dark matter is heated up by the decaying dark energy.
This kind of the enhancement of $\theta_m$ might significantly increase the velocity of dark matter.
To realize this effect, we have to further investigate  
physics at the scale of the dark matter halo, at which the linear perturbation theory is no longer valid, 
and the non-perturbative calculation is needed.

\section*{Acknowledgments}
This work was partially supported by National Center for Theoretical Sciences, National Science Council (NSC-101-2112-M-007-006-MY3), 
MoST (MoST-104-2112-M-007-003-MY3), and NSFC (11547008).

\end{document}